\documentclass[%
 aip,
 amsmath,amssymb,
 reprint,%
]{revtex4-1}

\usepackage{graphicx}
\usepackage{dcolumn}
\usepackage{bm}

\usepackage[utf8]{inputenc}
\usepackage[T1]{fontenc}
\usepackage{mathptmx}
\usepackage{etoolbox}
\usepackage{siunitx}
\usepackage{svg}
\usepackage{amsmath}

\definecolor{Myblue}{cmyk}{0.71,0.53,0,0.12}
\newcommand{\highlightchanges}[1]{#1}

\makeatletter
\def\@email#1#2{%
 \endgroup
 \patchcmd{\titleblock@produce}
 {\frontmatter@RRAPformat}
 {\frontmatter@RRAPformat{\produce@RRAP{*#1\href{mailto:#2}{#2}}}\frontmatter@RRAPformat}
 {}{}
}%
\makeatother

\begin{document}

\preprint{AIP/123-QED}

\title{Sideband Injection Locking in Microresonator Frequency Combs}
\author{Thibault Wildi}
\affiliation{Deutsches Elektronen-Synchrotron DESY, Notkestr. 85, 22607 Hamburg, Germany}
\author{Alexander Ulanov}%
\affiliation{Deutsches Elektronen-Synchrotron DESY, Notkestr. 85, 22607 Hamburg, Germany}
\author{Nicolas Englebert}%
\affiliation{OPERA-\textit{Photonics}, Universit\'e libre de Bruxelles (U.L.B.), 50~Avenue F. D. Roosevelt, CP 194/5, B-1050 Brussels, Belgium}
\author{Thibault Voumard}
\affiliation{Deutsches Elektronen-Synchrotron DESY, Notkestr. 85, 22607 Hamburg, Germany}
\author{Tobias Herr}
\affiliation{Deutsches Elektronen-Synchrotron DESY, Notkestr. 85, 22607 Hamburg, Germany}
\affiliation{Physics Department, Universität Hamburg UHH, Luruper Chaussee 149, 22761 Hamburg, Germany}%
\email{tobias.herr@desy.de}

\begin{abstract} 
Frequency combs from continuous-wave-driven Kerr-nonlinear microresonators have evolved into a key photonic technology with applications from optical communication to precision spectroscopy. Essential to many of these applications is the control of the comb’s defining parameters, i.e., \highlightchanges{carrier-envelope} offset frequency and repetition rate. 
An elegant and all-optical approach to controlling both degrees of freedom is the suitable injection of a secondary continuous-wave laser into the resonator onto which one of the comb lines locks. Here, we study experimentally such sideband injection locking in microresonator soliton combs across a wide optical bandwidth and derive analytic \highlightchanges{scaling laws} for the locking range and repetition rate control. As an application example, we demonstrate optical frequency division and repetition rate phase-noise reduction to three orders of magnitude below the noise of a free-running system.  
\highlightchanges{The presented results can guide the design of sideband injection-locked, parametrically generated frequency combs with opportunities for low-noise microwave generation, compact optical clocks with simplified locking schemes and more generally, all-optically stabilized frequency combs from Kerr-nonlinear resonators.}
\end{abstract}
\maketitle

\section{Introduction}
Continuous-wave (CW) coherently-driven Kerr-nonlinear resonators can create temporally structured waveforms that circulate stably without changing their temporal or spectral intensity profile. The out-coupled optical signal is periodic with the resonator roundtrip time $T_\mathrm{rep}$ and corresponds to an optical frequency comb\cite{delhaye:2007, kippenberg:2011, pasquazi:2018, fortier:2019, diddams:2020}, i.e. a large set of laser frequencies spaced by the repetition rate $f_\mathrm{rep}=T_\mathrm{rep}^{-1}$. 
One important class of such stable waveforms are CW-driven dissipative Kerr-solitons (DKSs), which have been observed in fiber-loops\cite{leo:2010}, traveling- and standing-wave microresonators\cite{herr:2014b, wildi:2023} and free-space cavities\cite{lilienfein:2019}. In microresonators these soliton microcombs\cite{kippenberg:2018} provide access to low-noise frequency combs with ultra-high repetition rates up to THz frequencies, enabling novel applications in diverse fields including optical communication\cite{marin-palomo:2017, jorgensen:2022}, ranging\cite{trocha:2018, suh:2018, riemensberger:2020}, astronomy\cite{suh:2019, obrzud:2019a}, spectroscopy\cite{suh:2016}, microwave photonics\cite{liang:2015a, lucas:2020}, and all-optical convolutional neural networks\cite{feldmann:2021}.

\highlightchanges{In a CW-driven microresonator, the} comb's frequency components are defined by $f_\mu = f_\mathrm{p} + \mu f_\mathrm{rep}$, where $f_\mathrm{p}$ denotes the frequency \highlightchanges{of the central comb line} and $\mu$ is the index of the comb line with respect to the central line ($\mu$ is also used to index the resonances supporting the respective comb lines). 
For many applications\highlightchanges{\cite{fortier:2019, diddams:2020}}, it is essential to control both degrees of freedom in the generated frequency comb spectra, i.e. the repetition rate $f_\mathrm{rep}$ and the \highlightchanges{central} frequency $f_\mathrm{p}$ \highlightchanges{(which together define the comb's carrier-envelope offset frequency)}. Conveniently, for Kerr-resonator based combs, \highlightchanges{$f_\mathrm{p}$} is defined by the pump laser frequency $f_\mathrm{p} = \omega_\mathrm{p}/(2\pi)$. However, the repetition rate $f_\mathrm{rep}$ depends on the resonator and is subject to fundamental quantum mechanical as well as environmental fluctuations. 

\begin{figure}[ht!]
	\includegraphics[width=\columnwidth]{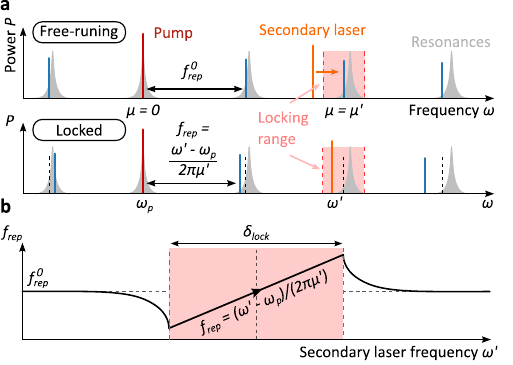}
	\caption{\textbf{Principles of sideband injection locking.}
		\textbf{a}, In a free-running comb, the central comb line is defined by the pump laser around which equidistant comb lines, spaced by the free-running repetition rate $f_\mathrm{rep}^{0}$, are formed. If a secondary injection laser of frequency $\omega'$ is brought close to one of the comb lines (within injection locking range), then the comb locks to the injecting laser, modifying the repetition rate as indicated.
		\textbf{b} Outside the locking range, $f_\mathrm{rep}=f_\mathrm{rep}^{0}$ is unaffected by the secondary laser. Inside the locking range, it follows a characteristic tuning behavior with a linear dependence on the injecting laser frequency $\omega'$.
		\label{fig:concept}
	}
\end{figure}

A particularly attractive and all-optical approach to controlling $f_\mathrm{rep}$ is the injection of a secondary CW laser of frequency $\omega'$ into the resonator\highlightchanges{, demonstrated numerically\cite{taheri:2017} and experimentally\cite{lu:2021}}. If $\omega'$ is sufficiently close to one of the free-running comb lines (sidebands) $f_\mu \approx \omega'/(2\pi)$, i.e., within \textit{locking range}, the comb will lock onto the secondary laser, so that $f_\mu \rightarrow \omega'/(2\pi)$. The repetition rate is then $f_\mathrm{rep} = (\omega_\mathrm{p}-\omega')/(2\pi\mu')$, with $\mu'$ denoting the index of the closest resonance to which the secondary laser couples, cf. Fig.~\ref{fig:concept}a. This frequency division\cite{fortier:2011} of the frequency interval defined by the two CW lasers (as well as their relative frequency noise) by the integer $\mu'$ can give rise to a low-noise repetition rate $f_\mathrm{rep}$. 
In previous work, sideband injection locking has been leveraged across a large range of photonic systems, including for parametric seeding\cite{papp:2013, taheri:2015}, 
dichromatic pumping\cite{hansson:2014a}, 
optical trapping\cite{jang:2015, taheri:2017, erkintalo:2022},
synchronization of solitonic and non-solitonic combs\cite{jang:2018, kim:2021},
soliton crystals\cite{lu:2021}, soliton time crystals\cite{taheri:2022}, multi-color solitons\cite{moille:2022} and optical clockworks by injection into a DKS dispersive wave\cite{moille:2023}.
Related dynamics also govern the self-synchronization of comb states\cite{delhaye:2014, taheri:2017a}, the binding between solitons\cite{wang:2017a}, \highlightchanges{modified soliton dynamics in the presence of Raman-effect\cite{englebert:2023} and avoided mode-crossings\cite{taheri:2023}}, as well as the respective interplay between co-\cite{suzuki:2019} and counter-propagating solitons\cite{yang:2017, yang:2019, wang:2023} \highlightchanges{and multi-soliton state-switching\cite{taheri:2022a}}. Moreover, sideband injection locking is related to modulated and pulsed driving for broadband stabilized combs\cite{obrzud:2017, obrzud:2019a, anderson:2021}, as well as spectral purification and non-linear filtering of microwave signals\cite{weng:2019a, brasch:2019} via DKS.
Despite the significance of sideband injection locking, a broadband characterization and quantitative understanding of its dependence on the injecting laser are lacking, making the design and implementation of such systems challenging.

In this work, we study the dynamics of sideband injection locking with DKS combs. Our approach leverages 
high-resolution coherent spectroscopy of the microresonator under DKS operation, enabling precise mapping of locking dynamics \highlightchanges{across a large set of comb modes, including both the central region and wing of the comb.} 
We derive the sideband injection locking range's dependence on experimentally accessible parameters and find excellent agreement with the experimental observation\highlightchanges{ and with numeric simulation. Specifically, this includes the square dependence on the mode number, the square-root dependence on injection laser and DKS spectral power, as well as, the associated spectral shifts.
} 
In addition, we demonstrate \highlightchanges{experimentally} optical frequency division and repetition rate phase-noise reduction \highlightchanges{in a DKS state} to three orders of magnitude below the noise of a free-running system.

\section{Results}
\begin{figure}[b!]
	\includegraphics[width=\columnwidth]{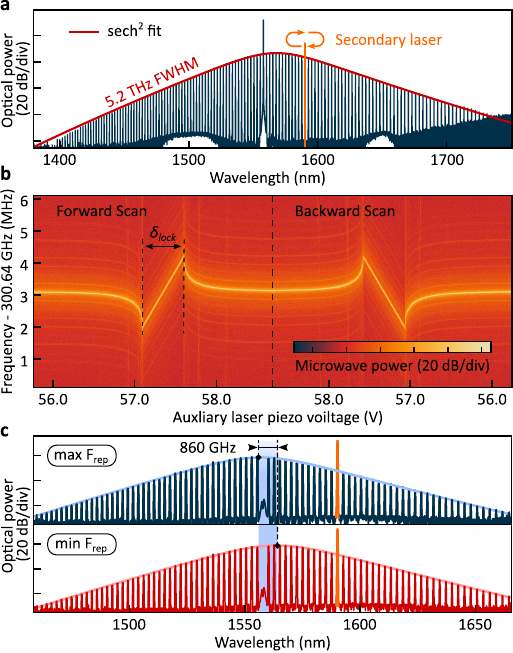}
	\caption{\textbf{Soliton sideband injection locking.} 
        \textbf{a}, Single DKS comb spectrum, following a $sech^2$ envelope, with a full-width-at-half-maximum (FWHM) of \SI{5.2}{\THz}, corresponding to a \SI{\sim 60}{\fs} pulse. The secondary laser is introduced in the spectral wing of the soliton and scanned across the \mbox{$\mu'\,$$^\mathrm{th}$} sideband.
        \textbf{b}, Repetition rate beatnote observed while the secondary laser is scanned across the \mbox{$\mu'\,$$^\mathrm{th}$} sideband. The locking bandwidth corresponds to the region of linear evolution of the repetition rate beatnote.
        \textbf{c}, Spectra of two sideband injection-locked DKS states from either end of the locking range, exhibiting a differential spectral shift of \SI{860}{\GHz}. Note that a filter blocks the central pump component $\omega_\mathrm{p}$.
	\label{fig:intro}
	}
\end{figure}
To first explore the sideband injection locking dynamics experimentally, we generate a single DKS state in a silicon nitride ring-microresonator. \highlightchanges{In the fundamental TE modes, the resonator is characterized by a quality factor of $Q\approx2$~million (linewidth $\kappa/(2\pi)\approx\SI{100}{\MHz}$), a free-spectral range (FSR) of $D_1/(2\pi) = \SI{300}{\GHz}$ and exhibits anomalous group velocity dispersion $D_2/(2\pi) = \SI{9.7}{\MHz}$ so that the resonance frequencies are well-described by $\omega_\mu=\omega_0 + \mu D_1 + \mu^2 \frac{D_2}{2}$ ($1.6 \times \SI{0.8}{\um^2}$ cross-section, \SI{76}{\um} radius). To achieve deterministic single soliton initiation, the microresonator's inner perimeter is weakly corrugated\cite{yu:2021, ulanov:2023}. The resonator is critically coupled and driven by a CW pump laser (\SI{\sim300}{\kHz} linewidth) with on-chip power of \SI{200}{\mW} at \SI{1557}{\nm} (pump frequency $\omega_\mathrm{p}/(2\pi)=\SI{192.5}{\THz}$)\cite{herr:2014b}.}
The generated DKS has a \SI{3}{\decibel} bandwidth of approximately \SI{5.2}{\THz} (cf. Fig.~\ref{fig:intro}a) corresponding to a bandwidth limited pulsed of \SI{\sim60}{\fs} duration. 
The soliton spectrum closely follows a $sech^2$ envelope and is free of dispersive waves or avoided mode crossings. The spectral center of the soliton does not coincide with the pump laser but is slightly shifted towards longer wavelengths due to the Raman self-frequency shift\cite{karpov:2016,yi:2016a}. 

\begin{figure*}[ht!]
	\includegraphics[width=\textwidth]{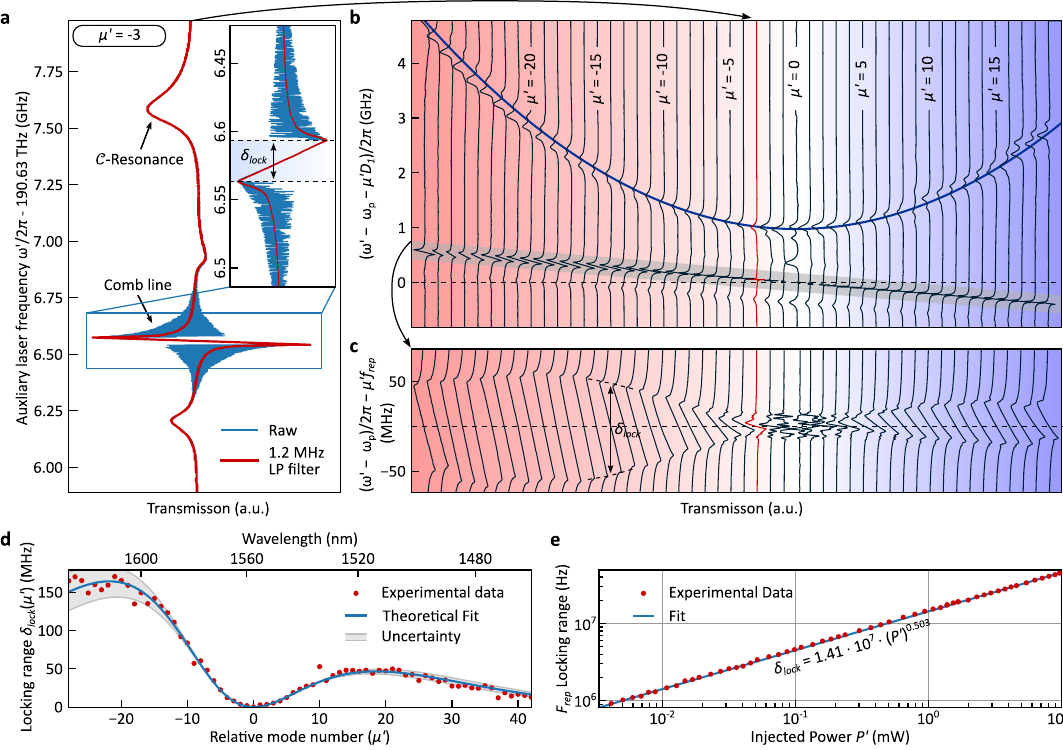}
	\caption{\textbf{DKS sideband injection locking dynamics.}
  \textbf{a},~Transmission obtained when the secondary laser frequency $\omega'$ is scanned in the vicinity of comb line $\mu' = -3$. The trace contains features indicating the position of the microresonator resonance frequency $\omega_{-3}/(2\pi)$ and of the soliton comb line frequency $f_{-3}$ as well as the sideband injection locking range (see main text for details). 
  \textbf{b},~Similar to \textbf{a} but for all $\mu'$ that can be reached by the scanning laser frequency $\omega'$. In this representation, the resonance frequencies form a quadratic integrated dispersion profile (due to anomalous dispersion) while the equidistant soliton microcomb lines (highlighted in gray and expanded in panel \textbf{b}) form a straight line, enabling retrieval of pump laser detuning and microcomb repetition rate (see main text for details). 
		\textbf{c},~Zoom into \textbf{b}, focusing on the vicinity of the comb lines. The spectral dependence of the locking range can be observed (cf. panel \textbf{a} and see main text for details).
	 \textbf{d},~Locking range as a function of the relative mode number $\mu'$. The measured data closely follows the predicted scaling (cf. main text). \highlightchanges{The grey area indicates the uncertainty we expect from 10\% detuning fluctuations during the recording procedure.}
 		\textbf{e},~Locking range \highlightchanges{in terms of the repetition rate $f_{rep}$} for $\mu' = -13$ as a function of secondary pump power (estimated on-chip power). \highlightchanges{Analogous to \textbf{d}, the uncertainty is approx. $\pm 4\%$.}
 }
	\label{fig:wide-scan}
\end{figure*}

A secondary CW laser \highlightchanges{(\SI{\sim300}{\kHz} linewidth)}, tunable both in power and frequency (and not phase-locked in any way to the first CW laser), is then combined with the pump laser upstream of the microresonator and scanned across the \mbox{$\mu'\,$$^\mathrm{th}$} sideband of the soliton microcomb, as illustrated in Fig.~\ref{fig:intro}a. 
The spectrogram of the repetition rate signal recorded during this process is shown in Fig.~\ref{fig:intro}b, for $\mu' = -13$, and exhibits the canonical signature of locking oscillators\cite{adler:1946} (cf. Supplementary Information (SI), Section~\ref{SI:measure_frep} for details on the measurement of $f_\mathrm{rep}$). Specifically, the soliton repetition rate $f_\mathrm{rep}$ is observed to depend linearly on the auxiliary laser frequency $\omega'$ over a locking range $\delta_\mathrm{lock}$ following $f_\mathrm{rep} = \frac{1}{2\pi}\frac{\omega_\mathrm{p} - \omega'}{\mu'}$. Within $\delta_\mathrm{lock}$, the soliton comb latches onto the auxiliary laser, such that the frequency of the comb line with index $\mu'$ is equal to the secondary laser frequency. The locking behavior is found to be symmetric with respect to the scanning direction, and no hysteresis is observed. Figure~\ref{fig:intro}c shows the optical spectra of two sideband injection-locked DKS states, with the secondary laser positioned close to the respective boundaries of the locking range. A marked shift of the spectrum of \SI{860}{\GHz} is visible when going from one state to the other. \highlightchanges{As we will discuss below and in the SI, Section~\ref{SI:waveform_velocity}, the spectral shift in the presence of non-zero group velocity dispersion modifies the soliton's group velocity and provides a mechanism for the DKS to adapt to the repetition rate imposed by the driving lasers. }

Having identified characteristic features of sideband injection locking in our system, we systematically study the injection locking range and its dependence on the mode number $\mu'$ to which the secondary laser is coupled. To this end, a frequency comb calibrated scan\cite{delhaye:2009} of the secondary laser's frequency $\omega'$ across many DKS lines is performed. The power transmitted through the resonator coupling waveguide is simultaneously recorded. It contains the $\omega'$-dependent transmission of the secondary laser as well as the laser's heterodyne mixing signal with the DKS comb, which permits retrieving the locking range $\delta_\mathrm{lock}$.

Figure~\ref{fig:wide-scan}a shows an example of the recorded transmission signal where the scanning laser's frequency $\omega'$ is in the vicinity of the comb line with index $\mu'=-3$. When the laser frequency $\omega'$ is sufficiently close to the DKS comb line, the heterodyne oscillations (blue trace) can be sampled; when $\omega'$ is within the locking range $\delta_\mathrm{lock}$, the heterodyne oscillations vanish, and a linear slope is visible, indicating the changing phase between the comb line and the secondary laser across the injection locking range. In addition to the heterodyne signal between the comb line and laser, a characteristic resonance feature, the so-called $C$-resonance\cite{matsko:2015, lucas:2017}, representing (approximately) the resonance frequency $\omega_\mu$ is observed. 

The set of equivalent traces for all comb lines $\mu'$ in the range of the secondary (scanning) laser is presented in Fig.~\ref{fig:wide-scan}b as a horizontal stack. For plotting these segments on a joint vertical axis, $\omega_\mathrm{p} + \mu' D_1$ has been subtracted from $\omega'$.
In this representation, the parabolic curve (blue line in Fig.~\ref{fig:wide-scan}b) connecting the $C$-resonances signifies the anomalous dispersion of the resonator modes $\omega_\mu$. In contrast, the equidistant comb lines form a straight feature (grey highlight), of which a magnified view is presented in Fig.~\ref{fig:wide-scan}c. \highlightchanges{Due to the Raman self-frequency shift, the free-running repetition rate of the DKS comb $f_\mathrm{rep}^0$ is smaller than the cavity's FSR $D_1/(2\pi)$, resulting in the negative tilt of the line.} 
Here, to obtain a horizontal arrangement of the features, $\omega_\mathrm{p} + \mu' 2\pi f_\mathrm{rep}^0$ has been subtracted from $\omega'$.
The locking range $\delta_\mathrm{lock}$ corresponds to the vertical extent of the characteristic locking feature in Fig.~\ref{fig:wide-scan}c. Its value is plotted as a function of the mode number in Figure~\ref{fig:wide-scan}d, revealing a strong mode number dependence of the locking range with local maxima (almost) symmetrically on either side of the central mode. 
The asymmetry in the locking range with respect to $\mu' =0 $ (with a larger locking range observed for negative values of $\mu'$) coincides with the Raman self-frequency shift of the soliton spectrum (higher spectral intensity for negative $\mu$). Next, we keep $\mu'$ fixed and measure the dependence of $\delta_\mathrm{lock}$ on the power of the injecting laser $P'$. As shown in Fig.~\ref{fig:wide-scan}e, we observe an almost perfect square-root scaling $\delta_\mathrm{lock} \propto \sqrt{P'}$, revealing the proportionality of the locking range to the strength of the injected field.

The observed \highlightchanges{scaling of the locking range} may be understood in both the time and frequency domain. 
In the time domain, the beating between the two driving lasers creates a modulated background field inside the resonator, forming an optical lattice trap for DKS pulses\cite{jang:2015, taheri:2017}. Here, to derive the injection locking range $\delta_\mathrm{lock}$, we extend the approach proposed by Taheri et al.\cite{taheri:2015}, which is based on the momentum $p=\sum_\mu \mu |a_\mu|^2 = \bar{\mu}\sum_\mu |a_\mu|^2$ of the waveform (in a co-moving frame), where $a_\mu$ is the complex field amplitude in the mode with index $\mu$, normalized such that $|a_\mu|^2$ scales with the photon number and $\bar{\mu}$ the \emph{photonic center of mass} in mode number/photon momentum space. As we show in the SI, Section~\ref{SI:locking_time_domain}, the secondary driving laser modifies the waveform's momentum, thereby its propagation speed and repetition rate. For the locking range of the secondary laser, we find
\begin{equation}
 \delta_\mathrm{lock} = \frac{2}{\pi} \mu'^2 \eta D_2 \frac{\sqrt{P' P_{\mu'}}}{\sum_\mu P_\mu} \frac{\omega_\mathrm{p}}{\omega_{\mu'}},
 \label{eq:locking_range}
\end{equation}
and for the repetition rate tuning range 
\begin{equation}
  \delta f_\mathrm{rep}= \delta_\mathrm{lock}/ | \mu' |,
  \label{eq:frep_range}
\end{equation}
where $\eta$ is the coupling ratio, and the $P_\mu$ refer to the spectral power levels of the comb lines with index $\mu$ measured outside the resonator. The spectral shift of the spectrum in units of mode number $\mu$ is $2\pi \delta f_\mathrm{rep}/D_2$. \highlightchanges{In the SI, Section~\ref{SI:Injection_ratio}, we recast Eq.~\ref{eq:frep_range} in terms of the injection ratio $\mathrm{IR}=P'/P_{\mu'}$ to enable comparison with CW laser injection locking\cite{hadley:1986}.}
\highlightchanges{The results in Eqs.~\ref{eq:frep_range} and \ref{eq:locking_range} may also be obtained in a frequency domain picture \highlightchanges{(see SI, Section~\ref{SI:locking_frequency_domain})}, realizing that the waveform's momentum is invariant under Kerr-nonlinear interaction (neglecting the Raman effect) and hence entirely defined by the driving lasers and the rate with which they inject photons of specific momentum into the cavity (balancing the cavity loss). If only the main pump laser is present, then $\bar{\mu}=0$. However, in an injection-locked state, depending on phase, the secondary pump laser can coherently inject (extract) photons from the resonator, shifting $\bar{\mu}$ towards (away from) $\mu'$. This is equivalent to a spectral translation of the intracavity field, consistent with the experimental evidence in Fig.~\ref{fig:intro}c.} 

\highlightchanges{To verify the validity of Eq.~\ref{eq:locking_range} and \ref{eq:frep_range}, we perform numeric simulation (SI, Section~\ref{SI:simulation}) based on the Lugiato-Lefever Equation (LLE) (see SI, Section~\ref{SI:definitions}). We find excellent agreement between the analytic model and the simulated locking range.
We note that Eq.~\ref{eq:locking_range} and \ref{eq:frep_range} are derived in the limit of low injection power, which we assume is the most relevant case. For large injection power, the spectrum may shift substantially and consequently affect the values of $P_\mu$. Interestingly, while this effect leads to an asymmetric locking range, the extent of the locking range is only weakly affected as long as the spectrum can locally be approximated by a linear function across a spectral width comparable to the shift. 
Injection into a sharp spectral feature (dispersive wave) is studied by Moille et al.\cite{moille:2023}
}

\highlightchanges{
The values of $P_\mu$ do not generally follow a simple analytic expression and can be influenced by the Raman effect and higher-order dispersion. While our derivation accounts for the values of $P_\mu$ (e.g., for the Raman effect $a_\mu$ and $P_\mu$ are increased (reduced) for $\mu$ below (above) $\mu = 0$), it does not include a physical description for Raman- or higher-order dispersion effects; these effects may further modify the locking range.
}
\highlightchanges{Taking into account the spectral envelope of the DKS pulse as well as the power of the injecting laser (which is not perfectly constant over its scan bandwidth), we fit the scaling $\delta_\mathrm{lock}\propto \mu'^2 \sqrt{P'P_{\mu'}}$ to the measured locking range in Fig.~\ref{fig:wide-scan}d, where we assume $P_{\mu'}$ to follow an offset (Raman-shifted) sech$^2$-function. The fit and the measured data are in excellent agreement, supporting our analysis and suggesting that the Raman shift does not significantly change the scaling behavior. Note that the effect of the last factor in Eq.~\ref{eq:locking_range} is marginal, and the asymmetry in the locking range is due to the impact of the Raman effect on $P_{\mu}$}.
\highlightchanges{It is worth emphasizing that our analysis did not assume the intracavity waveform to be a DKS state and we expect that the analytic approach can in principle also be applied to other stable waveforms, including those in normal dispersion combs \cite{xue:2016, kim:2021}. Indeed, as we show numerically in the SI, Section~\ref{SI:simulation}, sideband-injection locking is also possible for normal dispersion combs. Here, in contrast to a DKS, sideband laser injection is found to have a strong impact on the spectral shape (not only spectral shift). Therefore, although the underlying mechanism is the same as in DKS combs, Eq.~\ref{eq:locking_range} and Eq.~\ref{eq:frep_range} do not generally apply  (in the derivation, it is assumed that the spectrum does not change substantially).
}

\begin{figure}[ht]
	\includegraphics[width=\columnwidth]{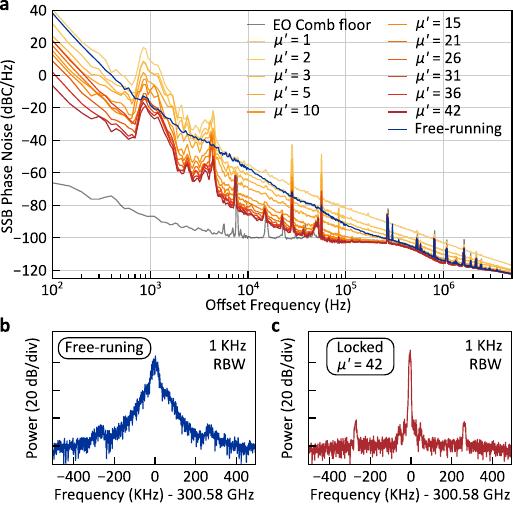}
	\caption{\textbf{Optical frequency division.}
		\textbf{a}, Repetition rate phase noise of the soliton microcomb in the free-running and locked states, with values of $\mu'$ ranging from 1 to 42. At higher offset frequencies (>\SI{100}{\kHz}), the phase noise of the electro-optic modulation used to down-mix the \SI{300}{\GHz} repetition rate signal to detectable frequencies (cf. SI) limits the measurement. 
		\textbf{b}, Repetition rate beat note recorded in the free-running state.
 		\textbf{c}, Repetition rate beatnote recorded in the locked state ($\mu' = 42$). The sidebands at approx. \SI{\pm300}{\kHz} are an artifact of the electro-optic modulation-based repetition rate detection scheme.
		\label{fig:pn}
	}
\end{figure}

Finally, as an example application of sideband injection locking, we demonstrate optical frequency division, similar to previous work \highlightchanges{\cite{moille:2023}}, and measure the noise reduction in $f_\mathrm{rep}$ (Fig.~\ref{fig:pn}a). With a growing separation between the two driving lasers (increasing $\mu'$), the phase noise is lowered by a factor of $\mu'^2$, resulting in a phase noise reduction of more than 3 orders of magnitude (with respect to the free-running case) when injecting the secondary laser into the mode with index $\mu'=42$ (limited by the tuning range of the secondary laser), and this without any form of stabilization of either the pump or secondary laser. Fig.~\ref{fig:pn}b and c compare the repetition rate beatnote of the free-running and injection-locked cases. 

\section{Conclusion}
In conclusion, we have presented an experimental 
and analytic study of sideband injection locking in DKS microcombs. The presented results reveal the dependence of the locking range on the intracavity spectrum and on the injecting secondary laser, with an excellent agreement between experiment and theory. While our experiments focus on the important class of DKS states, we emphasize that the theoretical framework from which we derive the presented scaling laws is not restricted to DKSs \highlightchanges{and may potentially be} transferred to other stable waveforms. Our results provide a solid basis for the design of sideband injection-locked, parametrically generated Kerr-frequency combs and may, in the future, enable new approaches to low-noise microwave generation, compact optical clocks with simplified locking schemes, and more generally, stabilized low-noise frequency comb sources from Kerr-nonlinear resonators.

\section*{Data Availability Statement}

The data supporting this study's findings are available from the corresponding author upon reasonable request.

\section*{Funding}
This project has received funding from the European Research Council (ERC) under the EU’s Horizon 2020 research and innovation program (grant agreement No 853564) and through the Helmholtz Young Investigators Group VH-NG-1404; the work was supported through the Maxwell computational resources operated at DESY.

\section*{References}

\bibliography{bibliography}

\onecolumngrid
\appendix

\newpage 
\section*{Supplementary Information}

\highlightchanges{
\subsection{Locking range equation in terms of the injection ratio}
\label{SI:Injection_ratio}
Eqs.~\ref{eq:locking_range} and \ref{eq:frep_range} can be recast in terms of the injection ratio IR=$P'/P_{\mu'}$ and read:
\begin{equation}
 \delta_\mathrm{lock} = \frac{2}{\pi} \mu'^2 \eta D_2 \frac{P_{\mu'}}{\sum_\mu P_\mu} \frac{\omega_\mathrm{p}}{\omega_{\mu'}} \sqrt{\mathrm{IR}} \propto \sqrt{\mathrm{IR}} 
\end{equation}
\begin{equation}
  \delta f_\mathrm{rep}= \frac{2}{\pi} |\mu'| \eta D_2 \frac{P_{\mu'}}{\sum_\mu P_\mu} \frac{\omega_\mathrm{p}}{\omega_{\mu'}} \sqrt{\mathrm{IR}} \propto \sqrt{\mathrm{IR}} 
\end{equation}
}

\subsection{Measuring the comb's repetition rate}
\label{SI:measure_frep}
\begin{figure*}[ht]
	\includegraphics[width=\textwidth]{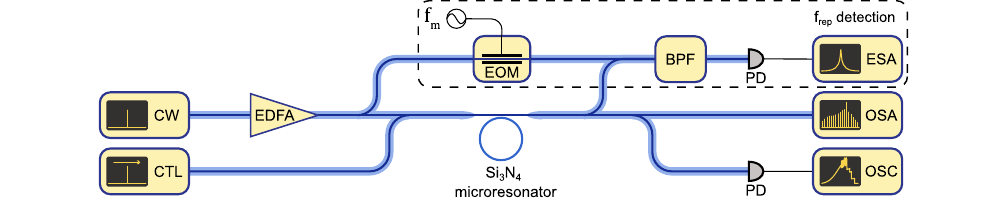}
	\caption{\highlightchanges{\textbf{Experimental setup.} A single-DKS state is generated inside a silicon nitride microring resonator using approximately \SI{200}{\mW} of pump power. A secondary continuously tunable laser is combined with the pump light before the cavity in order to investigate the sideband injection locking dynamics. In order to monitor the \SI{\sim 300}{\GHz} DKS repetition rate, we use a hybrid electro-optic detection scheme, similar to\cite{delhaye:2012} (see. supplementary information, section 1). EDFA: erbium-doped fiber amplifier; CW: continuous wave laser; CTL: continuously tunable laser; PD: photodetector; OSC: oscilloscope; ESA: electrical spectrum analyzer; OSA: optical spectrum analyzer; EOM: electro-optic modulator; BPF: band-pass filter.}
		\label{fig:setup}
	}
\end{figure*}

The soliton comb's repetition rate, too high for direct detection, is measured by splitting off a fraction of the pump light and phase-modulating it with frequency $f_\mathrm{m} = \SI{17.68}{\GHz}$, creating an electro-optic (EO) comb spanning \SI{\sim600}{\GHz} which is then combined with the DKS light \highlightchanges{(Fig.~\ref{fig:setup})}. 
A bandpass filter extracts the $17^\mathrm{th}$ line of the EO comb and the first sideband of the DKS comb, resulting in a beatnote at a frequency $f_\mathrm{s} = f_\mathrm{rep} - 17\, f_\mathrm{m}$ from which the soliton repetition rate $f_\mathrm{rep}$ can be recovered, similar to\cite{delhaye:2012}.

\subsection{Analytic description of sideband injection locking}

\subsubsection{Definitions}
\label{SI:definitions}
We start from the dimensionless form of the Lugiato-Lefever equation (LLE)\cite{lugiato:1987, chembo:2013}\highlightchanges{, describing the dynamics in a frame moving with the (angular) velocity $d_1$ (the angular interval $[0,2\pi]$ corresponds to one resonator round-trip):}
 \begin{equation}
 \begin{split}
  \frac{\partial\Psi}{\partial\tau} = & - (1 + i\zeta_0) \Psi + i |\Psi|^2\Psi +i \sum_{n=2}^{N\ge2} (-i)^{n}\frac{d_n}{n!}\frac{\partial^n \Psi}{\partial\theta^n} 
   + f,
 \end{split} 
\end{equation}
where $\theta$ is the azimuthal coordinate, $\tau = \kappa t / 2 $ denotes the normalized time ($\kappa$ being the cavity decay rate/total linewidth), $\Psi(\theta, \tau)$ is the waveform, $\zeta_0 = 2(\omega_0 - \omega_\mathrm{p})/\kappa$ the normalized pump detuning, $d_n = 2 D_n / \kappa$ the normalized dispersion coefficients and \highlightchanges{$f= s \sqrt{8 \eta g / \kappa^2}$ a pump field where $|s|^2 = \frac{P_\mathrm{p}}{\hbar\omega_\mathrm{p}}$ is the pump photon flux. The microresonator's coupling rate to the bus waveguide is $\kappa_\mathrm{ext}$ and $\eta = \kappa_\mathrm{ext} / \kappa$ is the microresonator's coupling coefficient}. Let $a_\mu\highlightchanges{(\tau)}$ be the normalized complex mode amplitudes such that:
\begin{align}
	\Psi(\theta\highlightchanges{,\tau}) & = \sum_\mu a_\mu\highlightchanges{(\tau)} e^{i\mu\theta}, \label{eq:fourrier1} \\
	a_\mu\highlightchanges{(\tau)} & = \frac{1}{2\pi} \int_0^{2\pi}\Psi(\theta\highlightchanges{,\tau})e^{-i\mu\theta} \mathrm{d} \label{eq:fourrier2}\theta, 
\end{align}
where $\mu$ is the relative mode number and $a_\mu$ is related to the circulating intracavity power $P_\mu$ via $ |a_\mu|^2 = \frac{2g}{\kappa} \frac{P_\mu}{\hbar \omega_\mu} \frac{2 \pi}{D_1} $. \highlightchanges{Here $g=\hbar\omega_0^2 c n_2 / (n_0^2 V_\mathrm{eff})$ is the nonlinear coupling coefficient , where $c$ is the speed of light, $n_0$ the refractive index, $n_2$ the nonlinear index, and $V_\mathrm{eff}$ the mode volume.}

The momentum $p$ of the intra-cavity field $\Psi(\theta \highlightchanges{, \tau})$ is defined as:
\begin{align}   
	p & := \frac{1}{2} \frac{1}{2\pi} \int_0^{2\pi} \highlightchanges{\left[ \Psi^* \left ( i\frac{\partial \Psi }{\partial\theta}\right ) + c.c. \right]} \:\mathrm{d} \theta \\
  \label{eq:momentum_1} 
	 & = \sum_\mu \mu | a_\mu |^2 \\
 & = \bar{\mu} \sum_\mu |a_\mu|^2 \highlightchanges{ = \bar{\mu} N }
 \label{eq:momentum_2} 
\end{align}
where we used Eq.~\ref{eq:fourrier1}, $\bar{\mu}$ denotes the \emph{photonic center of mass} and \highlightchanges{$N=\sum_\mu |a_\mu|^2$ scales with the number of photons in the cavity.}

\subsubsection{Waveform velocity, spectral shift and repetition rate}
\label{SI:waveform_velocity}
\highlightchanges{We assume a waveform $\Psi$ with a stable non-flat temporal intensity profile inside the resonator, i.e. the shape of the intensity profile does not change. The Waveform $\Psi$ may not be static (in the frame moving with angular velocity $d_1$), but move with an additional angular velocity component $\dot{\theta}$, so that
\begin{equation}
 |\Psi(\theta, 0)|^2 =|\Psi(\theta + \dot{\theta} \tau,\tau)|^2 \,.
\end{equation}
In a new coordinate frame ($\theta'=\theta-\dot{\theta}\tau$, $\tau'=\tau$, $\Psi'(\theta', \tau')=\Psi(\theta,\tau)$) that is co-moving with the intensity envelope, $|\Psi'|^2$ will be static so that $\partial |\Psi'|^2 /\partial\tau'=0$. For the derivatives we find
\begin{equation}
 \frac{\partial \Psi'}{\partial\theta'} = \frac{\partial \Psi}{\partial \theta}\frac{\partial \theta}{\partial \theta'} + \frac{\partial \Psi}{\partial \tau}\frac{\partial \tau}{\partial \theta'} = \frac{\partial \Psi}{\partial \theta} 
\end{equation}
and
\begin{equation}
 \frac{\partial \Psi'}{\partial\tau'} = \frac{\partial \Psi}{\partial \theta}\frac{\partial \theta}{\partial \tau'} + \frac{\partial \Psi}{\partial \tau}\frac{\partial \tau}{\partial \tau'} = \dot{\theta} \frac{\partial \Psi}{\partial \theta} + \frac{\partial \Psi}{\partial \tau} 
\end{equation}
so that
\begin{align}
\label{eq:static_intensity}
\begin{split}
 0=\frac{\partial \left | \Psi' \right | ^2}{\partial \tau'} & = \frac{\partial \Psi'}{\partial \tau'} \Psi'^* + \Psi' \frac{\partial \Psi'^*}{\partial \tau'} \\
 & = \frac{\partial \Psi}{\partial \tau} \Psi^* + \dot{\theta} \frac{\partial \Psi}{\partial \theta} \Psi^* + \Psi \frac{\partial \Psi^*}{\partial \tau} + \dot{\theta} \Psi\frac{\partial \Psi^*}{\partial \theta} \\
 & = \left ( \frac{\partial \Psi}{\partial \tau} \Psi^* + \Psi \frac{\partial \Psi^*}{\partial \tau} \right ) + \dot{\theta} \left ( \frac{\partial \Psi}{\partial \theta} \Psi^* + \Psi\frac{\partial \Psi^*}{\partial \theta} \right ) \\
 & = \left ( \frac{\partial \Psi}{\partial \tau} \Psi^* + \Psi \frac{\partial \Psi^*}{\partial \tau} \right ) + \dot{\theta} \frac{\partial \left | \Psi \right|^2}{\partial \theta} \, .
\end{split}
\end{align}
By replacing the time derivatives with the right side of the LLE and only accounting for second-order dispersion ($d_n = 0, \forall n \ge 3$), one finds
\begin{equation}
\label{eq:derivative_from_LLE}
 \frac{\partial \Psi}{\partial \tau} \Psi^* + \Psi \frac{\partial \Psi^*}{\partial \tau} = -2 |\Psi|^2 + f \Psi^* + f^* \Psi + \frac{id_2}{2} \left ( \Psi \frac{\partial^2 \Psi^*}{\partial \theta^2} - \Psi^* \frac{\partial^2 \Psi}{\partial \theta^2} \right ).
\end{equation}
Under the assumption that the pump and loss term cancel in eq.~\ref{eq:derivative_from_LLE}, eq.~\ref{eq:static_intensity} becomes 
\begin{align}
\begin{split}
 \label{eq:towards_v_01}
 0 & = \frac{id_2}{2} \left ( \Psi \frac{\partial^2 \Psi^*}{\partial \theta^2} - \Psi^* \frac{\partial^2 \Psi}{\partial \theta^2} \right ) + \dot{\theta} \frac{\partial \left | \Psi \right|^2}{\partial \theta} \\
 & = \frac{id_2}{2} \frac{\partial}{\partial \theta} \left ( \Psi \frac{\partial\Psi^*}{\partial \theta} - \Psi^* \frac{\partial \Psi}{\partial \theta} \right ) + \dot{\theta} \frac{\partial \left | \Psi \right|^2}{\partial \theta} \, .
\end{split}
\end{align}
Considering the indefinite integral $\int \mathrm{d}\theta$ results in
\begin{align}
\begin{split}
 \frac{id_2}{2} \left ( \Psi \frac{\partial\Psi^*}{\partial \theta} - \Psi^* \frac{\partial \Psi}{\partial \theta} \right ) + const = \dot{\theta} \left | \Psi \right|^2 \, .
\end{split}
\end{align}
In the following we take $const=0$, which in case of a non-zero value corresponds to a suitable (re-)definition of the moving frame. 
Next, integrating according to $\frac{1}{2\pi}\int_0^{2\pi} \mathrm{d}\theta$ and using Parseval's theorem $\sum_\mu |a_\mu|^2 = \frac{1}{2\pi}\int_0^{2\pi} |\Psi|^2 \mathrm{d}\theta = N$ gives
\begin{align}
\begin{split}
 \dot{\theta} = d_2\frac{p}{N} = d_2 \bar{\mu}
\end{split}
\end{align}

The change in the repetition rate $\delta f_\mathrm{rep}=\frac{\kappa}{2} \frac{\dot{\theta}}{(2\pi)} $, hence
\begin{equation}
 \label{eq:frep_p_mubar}
 \delta f_\mathrm{rep} = \frac{D_2}{2\pi}\frac{p}{N}= \frac{D_2}{2\pi}\bar{\mu} \, ,
\end{equation}
the change of the repetition rate is proportional to the shift of the photonic center of gravity $\bar{\mu}$ away from $\mu=0$.

}

\subsubsection{Sideband injection locking: Time domain description}
\label{SI:locking_time_domain}
In the case of a pump laser at frequency $\omega_\mathrm{p}$ and a secondary laser at frequency $\omega'$ close to $\omega_{\mu'}$, the pump field takes the form:
\begin{align}
 \label{eq:pump}
 f\rightarrow f(\theta, \tau) & = f_\mathrm{p} + f' e^{i \mu' \theta} e^{- i 2/\kappa(\omega' - \omega_\mathrm{p} - \mu' D_1)\tau} \\
 & = f_\mathrm{p} + f' e^{i \mu' \theta} e^{- i \Tilde{\zeta } \tau} 
\end{align}
where $ \Tilde{\zeta} = 2/\kappa(\omega' - \omega_\mathrm{p} - \mu' D_1)$ is a term describing the mismatch between the microresonator FSR and the grid defined by the pump and secondary lasers.

From ref.~\cite{taheri:2017}, Eq.~26, it follows that the force on the waveform $\Psi$ due to the presence of the secondary pump line is:
\begin{equation}
	\frac{\mathrm{d}p}{\mathrm{d}\tau} = -2 p + \mu'\left [f' e^{- i \Tilde{\zeta } \tau} \frac{1}{2\pi}\int_0^{2\pi}\Psi^* e^{i\mu'\theta} \mathrm{d} \theta + \text{.c.c} \right ]
\end{equation}
We recognize that the integral term is the Fourier transform of the intracavity field, such that: 
\begin{align}
	\frac{\mathrm{d}p}{\mathrm{d}\tau} & = -2 p + \mu' \left [f' e^{- i \Tilde{\zeta } \tau} a_{\mu'}^* + \text{c.c.} \right ]    \\
	\frac{\mathrm{d}p}{\mathrm{d}\tau} & = -2 p + 2 \mu' |f'| | a_{\mu'} | \cos(\Tilde{\zeta} \tau + \angle a_{\mu'} - \angle f') 
\end{align}
\highlightchanges{We assume that $\Psi$ is a stable waveform, moving at an angular velocity $\dot{\theta}$ within the LLE reference frame, which itself moves with $d_1$. In a suitable $(\theta',\tau')$-coordinate system, the waveform $\Psi'(\theta', \tau')$ is static (does not move), as described in SI, Section~\ref{SI:waveform_velocity}.
For the relation between the Fourier transforms of $\Psi$ and $\Psi'$ we find that $a_{\mu} = a'_{\mu} \exp{(-i \mu \dot{\theta} \tau)}$ and therefore $\angle a_{\mu'} = \angle a'_{\mu'} - \mu'\dot{\theta} \tau$. After substitution of the angle we find
\begin{equation}
 \label{eq:dp_dtau}
 \frac{\mathrm{d}p}{\mathrm{d}\tau} = -2 p + 2 \mu' |f'| | a_{\mu'} | \cos(\angle a'_{\mu'} - \angle f' + (\tilde{\zeta}- \mu' \dot{\theta}) \tau)
\end{equation}
We search for a steady state solution in which the momentum is constant $\mathrm{d}p / \mathrm{d}\tau = 0 $. As in a steady comb state $\frac{\partial a'_{\mu}}{\partial \tau}=0$ (and $f'$ does not depend on time), time independence is achieved when $\dot{\theta} = \tilde{\zeta} / \mu'$, i.e. the waveform must be moving at the velocity fixed by the pump and auxiliary laser detuning. Therefore, the momentum is purely a function of the relative phase between the secondary laser and the waveform's respective spectral component $\mu'$:}
\begin{equation}
	p = \mu' |f'| | a_{\mu'} | \cos(\angle a'_{\mu'} - \angle f')
\end{equation}
hence
\begin{equation}
	p \in [- \mu' |f'| | a_{\mu'} |; \mu' |f'| | a_{\mu'} |] \highlightchanges{\mathrm{\hspace{1cm}or \hspace{1cm}} \bar{\mu} \in \left[- \mu' \frac{|f'| | a_{\mu'}|}{N} ; \mu' \frac{|f'| | a_{\mu'}|}{N} \right]}
\end{equation}
\highlightchanges{
With Eq.~\ref{eq:frep_p_mubar}, this means that the repetition rate range in the injection-locked state is 
\begin{align}
 \begin{split}
  \delta f_\mathrm{rep} & = 2|\mu'|\frac{D_2}{2\pi}\frac{ |f'| | a_{\mu'} |}{N} \\
  & \approx 4 |\mu'| D_2 \frac{\sqrt{\kappa_\mathrm{ext} }}{\kappa} \frac{\sqrt{P' P_{\mu'}}}{\sum_\mu P_\mu} \frac{\omega_0}{\omega_{\mu'}} \\ 
 & = 4 |\mu'| \eta D_2 \frac{\sqrt{P' P_{\mu'}}}{\sum_\mu P_\mu} \frac{\omega_0}{\omega_{\mu'}} 
 \label{eq:frep_range_1}
 \end{split}
\end{align}

and the locking range is given by
\begin{align}
 \begin{split}
 \delta_\mathrm{lock} = | \mu' | f_\mathrm{rep} = 4 \mu'^2 \eta D_2 \frac{\sqrt{P' P_{\mu'}}}{\sum_\mu P_\mu} \frac{\omega_0}{\omega_{\mu'}} 
 \label{eq:locking_range_1}
 \end{split}
\end{align}
}
where the power levels $P'$ and $P_\mu$ are the power levels measured outside the
resonator. \highlightchanges{The approximation in the second line of Eq.~\ref{eq:locking_range_1} assumes that the mean frequency of the photons in the cavity is approximately $\omega_0$.}
Note that $P_{\mu'}$ and $P_0$ can readily be measured via a drop port; however, in a through-port configuration \highlightchanges{such as the one used in our experiment, it may be buried in residual pump light. For a smooth optical spectrum, their values may also be estimated based on neighboring comb lines for a smooth optical spectrum.}

\highlightchanges{
\subsubsection{Sideband injection locking: Frequency domain description}
\label{SI:locking_frequency_domain}
In the sideband injection-locked state, the nonlinear dynamics in the resonator may be described by the following set of coupled mode equations:
\begin{align}
\begin{split}
 \frac{\partial a'_{\mu}}{\partial \tau} = & - \left( 1 + i\frac{2}{\kappa}(\omega_{\mu} - \omega_\mathrm{p} - \mu \omega_\mathrm{R})\right)a'_{\mu} \\
 & + i\sum_{\alpha, \beta} a'_\alpha a'_\beta a^{\prime *}_{\alpha+\beta-\mu}\\
 & + \delta_{0\mu} \, f_\mathrm{p} + \delta_{\mu'\mu} \, f'
\end{split}
\end{align}
Here, the $a'_\mu$ represent the modes with the frequencies $\omega_\mathrm{p} + \mu \omega_\mathrm{R}=2\pi f_\mathrm{rep}$, where $\omega_\mathrm{R}$ is the actual repetition rate of the comb that may be different from $D_1$. Note that the $a'_\mu$ correspond to the $a'_\mu$ in SI, Section~\ref{SI:locking_time_domain}. In a steady comb state $\frac{\partial a'_{\mu}}{\partial \tau} =0$, so that a fixed phase relation between modes $a'_\mu$ and the external driving fields $f$ and $f'$ exists. We now consider only the field of the waveform, which we again denote with $a'_\mu$ for simplicity (Note that in a DKS, the separation of the DKS field at mode $\mu=0$ from that of the background is formally possible owing to their approximate phase shift of $\pi/2$). 
The rate at which photons are added to the waveform by the driving lasers is
\begin{align}
\begin{split}
 \left. \frac{\partial N_0}{\partial \tau} \right|_{f_\mathrm{p}} & = a^{\prime *}_0 \partial_t a'_0 + c.c. 
  = a^{\prime *}_0\, f_\mathrm{p} + c.c. = 2 |a'_0| |f_\mathrm{p}| \cos{(\angle a'_{0} - \angle f_\mathrm{p})}
\end{split}
\end{align}
for the main driving laser, and
\begin{align}
\begin{split}
 \left. \frac{\partial N_{\mu'}}{\partial \tau} \right|_{f'} & = a^{\prime *}_{\mu'} \partial_t a'_{\mu'} + c.c. 
  = a^{\prime *}_{\mu'}\, f' + c.c. = 2 |a'_{\mu'}| |f'| \cos{(\angle a'_{\mu'} - \angle f')}
\end{split}
\end{align}
for the secondary driving laser.
Photons are added or subtracted from the respective mode depending on the relative differential phase angles $\angle a'_{0} - \angle f_\mathrm{p}$ and $\angle a'_{\mu'} - \angle f'$. While $\angle a'_{0} - \angle f_\mathrm{p}\approx0$ for the main driving laser, $\angle a'_{\mu'} - \angle f'$ (for the secondary laser) can take all values from $0$ to $\pi$ during sideband injection locking. 
In the steady state the cavity losses are balanced by the lasers so that
\begin{align}
\begin{split}
 N = \sum_\mu N_\mu = \left. \frac{\partial N_0}{\partial \tau} \right|_{f_\mathrm{p}} + \left. \frac{\partial N_{\mu'}}{\partial \tau} \right|_{f'} 
\end{split}
\end{align}
The photonic center of gravity of the photons injected into the cavity is
\begin{align}
\begin{split}
 \bar{\mu} & = \frac{0\cdot \left. \frac{\partial N_0}{\partial \tau} \right|_{f_\mathrm{p}} + \mu' \cdot \left. \frac{\partial N_{\mu'}}{\partial \tau} \right|_{f'}}{ \left. \frac{\partial N_0}{\partial \tau} \right|_{f_\mathrm{p}} + \left. \frac{\partial N_{\mu'}}{\partial \tau} \right|_{f'}} 
 = \mu' \frac{ |a_{\mu'}|\, |f'| \cos{(\angle a'_{0} - \angle f')}}{\sum_\mu N_\mu} 
\end{split}
\end{align}
For clarity, we note that $\bar{\mu}$ does not change when photons are transferred through Kerr-nonlinear parametric processes from one mode to another mode. This is a consequence of (angular) momentum and photon number conservation (implying total mode number conservation) in Kerr-nonlinear parametric processes. Hence the photonic center of gravity of the injected photons (by main and secondary pump) is the same as the photonic center of gravity for the entire spectrum of the waveform. For the injection-locked state we find 
\begin{equation}
	\bar{\mu} \in \left[ -\mu' \frac{ |a_{\mu'}|\, |f'| }{\sum_\mu N_\mu}; \mu' \frac{ |a_{\mu'}|\, |f'| }{\sum_\mu N_\mu}\right]
\end{equation}
so that with Eq.~\ref{eq:frep_p_mubar} we obtain the same result as in the time domain description for $\delta f_\mathrm{rep}$ and $\delta_\mathrm{lock}$ (Eq.~\ref{eq:frep_range_1} and Eq.~\ref{eq:locking_range_1}).

\subsection{Numeric simulation of the sideband injection locking range}
\label{SI:simulation}

To complement our experimental and theoretical results, we run numerical simulations based on the coupled mode equation framework \cite{chembo:2010, hansson:2014a} (a frequency-domain implementation of the LLE). In order to observe the sideband injection locking dynamic, a single DKS is first initialized inside the cavity and numerically propagated. In the absence of a secondary laser, the soliton moves at the group velocity of the pump wavelength and appears fixed within the co-moving frame (Fig.~\ref{fig:simulation_1}a). The secondary laser $f'$ is then injected as per Eq.~\ref{eq:pump}, where $\Tilde{\zeta}$ controls the detuning of the secondary laser with respect to the free-running comb line $a_\mu$. For $ 0 < | \Tilde{\zeta}|  < \delta_\mathrm{lock} / \kappa$ (i.e., within the locking range), we observe that soliton moves at a constant velocity with respect to the co-moving frame (Fig.~\ref{fig:simulation_1}b). Beyond the locking range, the soliton is no longer phase-locked to the pump, which can readily be identified by tracking the relative phase between $a_{\mu'}$ and $f'$. We use this signature to identify the locking range from our simulations and compare it to our theoretical prediction from Eq.~\ref{eq:locking_range_1}; as can be seen in Fig.~\ref{fig:simualtion_2}, simulation and theory agree with striking fidelity.
\begin{figure*}[ht!]
	\includegraphics[width=\textwidth]{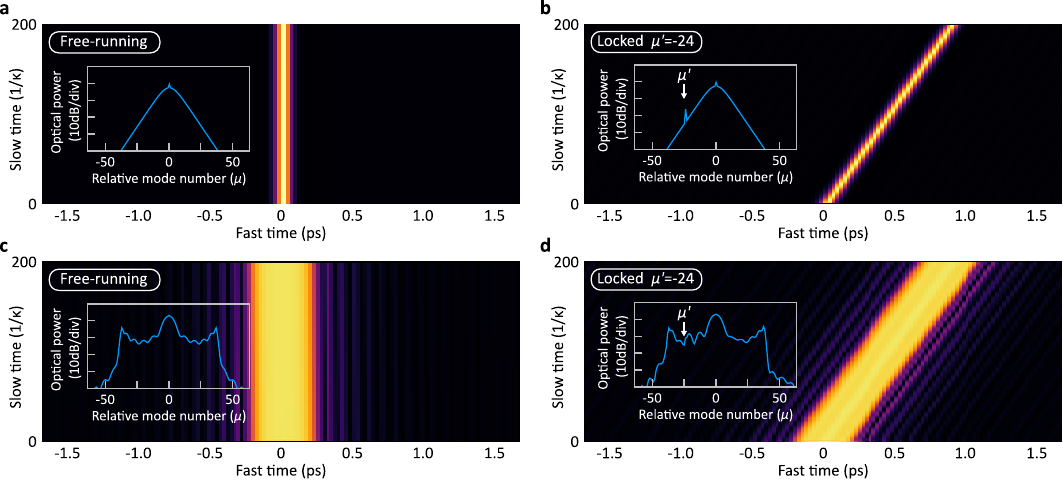}
	\caption{\highlightchanges{\textbf{Sideband injection locking simulation.} Evolution of the intracavity intensity profile in the free running (\textbf{a}, \textbf{c}) and locked states (\textbf{b}, \textbf{d}) for both DKS (\textbf{a}, \textbf{b}) and normal-dispersion combs (\textbf{c}, \textbf{d}). Inset: corresponding intracavity spectra.}
		\label{fig:simulation_1}
	}
\end{figure*}

We also study sideband injection locking dynamics inside normal-dispersion combs (Fig.~\ref{fig:simulation_1}c and d). Here as well, we observe the locking of the \emph{platicon} to the underlying modulation, although with a significant effect on its spectrum $P_\mu$ (see inset). This \emph{platicon}'s  spectrum lower robustness against external perturbation is not captured by our model which assumes a shifted but otherwise unchanged spectrum. Therefore, Eq.~\ref{eq:locking_range_1} does not generally apply, even though we expect our model to predict the locking range within a tolerance corresponding to the relative amplitude change of the corresponding comb line $a_{\mu'}$.
\begin{figure*}[ht!]
	\includegraphics[width=\textwidth]{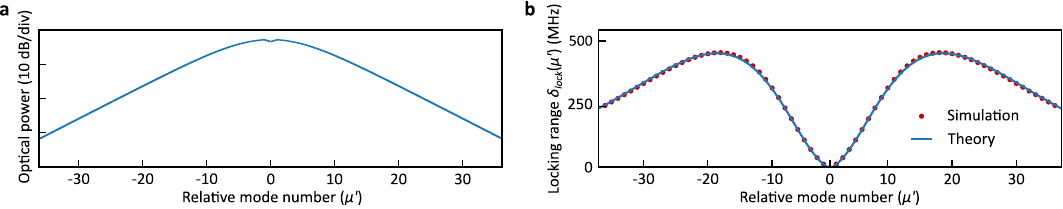}
	\caption{\highlightchanges{\textbf{Locking range simulations.} 
  \textbf{a}, Intracavity spectrum of the single DKS state used for the simulation of the sideband injection locking range.
    \textbf{b}, Simulated and theoretical (cf. Eq.~\ref{eq:locking_range_1}) locking range of the sideband-injection locking dynamic.
    }
		\label{fig:simualtion_2}
	}
\end{figure*}

\subsection{Effect of thermal resonance shifts}
\label{SI:thermal_Kerr_effects}
Across the sideband-injection locking range, the power in the comb state changes by small amounts due to the secondary laser, which may add or subtract energy from the resonator (cf. SI, Section~\ref{SI:locking_frequency_domain}). In consequence, the temperature of the resonator changes by small amounts, impacting via the thermo-refractive effect (and to a lesser extent by thermal expansion) the effective cavity length and hence $f_\mathrm{rep}$. This effect occurs on top of the sideband injection locking dynamics. In a typical microresonator, the full thermal shift of the driven resonance that can be observed prior to soliton formation is within 1 to 10~GHz. In a DKS state the coupled power is usually 1 to 10\%, implying a maximal thermal shift of 1~GHz. Assuming the secondary laser will change the power in the resonator by not more than 5\% (theoretical maximum for the highest pump power used in our manuscript), we expect a maximal thermal resonance shift of 50~MHz. Now, dividing by the absolute mode number of $>500$, we expect $\delta f_\mathrm{rep}<$100~kHz, which is two orders of magnitude below the effect resulting from sideband injection locking.
}

\end{document}